\documentclass[12pt]{article}
\usepackage{graphicx}

\usepackage{bm}
\usepackage{xcolor} 
\usepackage{amsmath,amsfonts,amssymb} 
\usepackage{tensor}
\usepackage{caption} 
\usepackage{subcaption}
\usepackage{dsfont}
\usepackage{braket}
\usepackage{dcolumn}
\newcommand{\LL}{\mathcal{L}}
\newcommand{\OO}{\mathcal{O}}

\newcommand\pubnumber{}
\newcommand\pubdate{\today}

\def\institute{DAMTP, University of Cambridge, Wilberforce Road, Cambridge, CB3 0WA, United Kingdom}
\def\support{\footnote{Work supported by the European Research Council under the European Union’s Horizon 2020 research and innovation Programme (grant agreement n.950246).}}

\def\Title#1{\begin{center} {\Large #1 } \end{center}}
\def\Author#1{\begin{center}{ \sc #1} \end{center}}
\def\Address#1{\begin{center}{ \it #1} \end{center}}

\newcommand\pubblock{\rightline{\begin{tabular}{l} \pubnumber\\
         \pubdate  \end{tabular}}}
\newenvironment{Abstract}{\begin{quotation}  }{\end{quotation}}
\newenvironment{Presented}{\begin{quotation} \begin{center} 
             PRESENTED AT\end{center}\bigskip 
      \begin{center}\begin{large}}{\end{large}\end{center} \end{quotation}}

\def\beq{\begin{equation}}
\def\eeq#1{\label{#1}\end{equation}}
\def\eeqn{\end{equation}}

\def\beqa{\begin{eqnarray}}
\def\eeqa#1{\label{#1}\end{eqnarray}}
\def\eeqan{\end{eqnarray}}

\let\bar=\overbar

\def\bra#1{\left\langle{ #1} \right|}
\def\ket#1{\left| {#1} \right\rangle}

\def\Dslash{\not{\hbox{\kern-4pt $D$}}}
\def\dslash{\not{\hbox{\kern-2pt $\del$}}}

\def\msb{{\bar{\ssstyle M \kern -1pt S}}}

\begin{document}
\begin{titlepage}
\pubblock

\vfill
\Title{Quantum SMEFT tomography: top quark pair}
\vfill
\Author{ Luca Mantani\support}
\Address{\institute}
\vfill
\begin{Abstract}
In recent years, an interest in the study of quantum information has grown within the high-energy particle physics community. The possibility to establish the presence of entanglement at particle colliders, such as the Large Hadron Collider (LHC) at CERN, is a novel and thrilling research direction, offering the opportunity to push quantum mechanics to its limits by using the heaviest fundamental particle: the top quark. With this in mind, we propose to use measurements of quantum entanglement to probe the behaviour of fundamental interactions and to search for New Physics at high energy within the Standard Model Effective Field Theory paradigm. Inspired by recent proposals to measure entanglement of top quark pairs produced at the LHC, we examine how the existence of new interactions between fundamental particles modify the Standard Model expectations. By performing analytical calculations, we unveil a non-trivial pattern of effects depending on the kinematical phase space configurations.
\end{Abstract}
\vfill
\begin{Presented}
$15^\mathrm{th}$ International Workshop on Top Quark Physics\\
Durham, UK, 4--9 September, 2022
\end{Presented}
\vfill
\end{titlepage}
\def\thefootnote{\fnsymbol{footnote}}
\setcounter{footnote}{0}

\section{Introduction}

The phenomenon of quantum entanglement has always been considered one of the most intriguing aspect of quantum mechanics. 
In particular, over the last decades, several studies have been performed proving the existence of entanglement and the violation 
of Bell's inequalities~\cite{PhysicsPhysiqueFizika.1.195}, ultimately helping us unveil and understand the complex world of
quantum phenomena.

In this direction, in recent years, a renovated interest in measuring entanglement at high energy has appeared. Specifically, 
top quark pairs look like ideal probes: they are vastly produced at the LHC and their spin information is retained by the decay products, allowing us to
fully reconstruct the spin density matrix of the system~\cite{Aoude:2022imd, Afik:2020onf, Severi:2021cnj,Afik:2022dgh,Afik:2022kwm, Aguilar-Saavedra:2022uye,Fabbrichesi:2021npl,Severi:2022qjy, Fabbrichesi:2022ovb}.

In this work~\cite{Aoude:2022imd}, we turn our attention to the study of the effects of heavy New Physics (NP) on entanglement 
patterns of top quark pair produced at the LHC. Specifically, with the aim of establishing how NP might induce a modification of the quantum correlations, we study the production of a top quark pair system within the Standard Model Effective Field Theory (SMEFT) framework, which provides a model independent formulation of new interactions when the energy scale associated to NP is well separated from the scale of the process.

\section{Spin density matrix}

The fundamental theoretical object to describe a quantum system is its density matrix. In particular, in the case of study, the 
spin density matrix of the top quark pair. The latter can be computed from the definition of the $R$-matrix, through the partonic scattering 
amplitude, i.e.
\begin{align}
    \label{eq:Rmatrix_Amplitudes}
    R^{I}_{\eta_1\eta_2,\zeta_1\zeta_2} &\equiv \frac{1}{N_a N_b}\sum_{\substack{\text{colors}\\\text{a,b spins}}} \hspace{-.5em} \mathcal{M}_{\eta_2 \zeta_2}^*\,  \mathcal{M}_{\eta_1 \zeta_1} \, ,
\end{align}
with $\mathcal{M}_{\eta\zeta} \equiv \langle t(k_1,\eta)\bar{t}(k_2,\zeta)|\mathcal{T}|a(p_1)b(p_2)\rangle$,
where $\mathcal{T}$ is the transition matrix element, $I=ab$ denotes the initial state, $N_{a,b}$ is the number of degrees of freedom of the respective initial state particles $a$ and $b$, $k_i$ ($p_i$) are the momenta of the final (initial) state particles, and $\eta$ ($\zeta$) are the (anti-)top spin indices.
Note that this matrix is similar to the cross-section, but with uncontracted final-state spin-indices. 

The $R$-matrix can be decomposed in terms of the spin operators~\cite{RevModPhys.55.855} in the following manner
\begin{equation}
\label{eq:FanoDecomposition}
R = \tilde{A}\, \mathds{1}_2\otimes \mathds{1}_2
+ \tilde{B}_i^+\sigma^i\otimes \mathds{1}_2 
+ \tilde{B}_i^- \mathds{1}_2 \otimes \sigma^i
+ \tilde{C}_{ij}\,\sigma^i\otimes \sigma^j \, ,
\end{equation}
where the $\tilde{C}_{ij}$ describe the spin correlations, the $\tilde{B}_i^{\pm}$ the net polarisation of the top quarks and
$\tilde{A}$ encodes information on the differential cross section
\begin{equation}
    \frac{\mathrm{d} \sigma}{\mathrm{d} \Omega}=\frac{\beta}{16 \pi^2\hat{s}} \tilde{A}(\hat{s}, \boldsymbol{k}) \, ,
\end{equation}
where $\bm{k}$ is the top quark direction, $\hat{s}$ the invariant mass of the top quark pair and $\beta = \sqrt{1-4m_t^2/\hat{s}}$ the velocity of the top quark in the center-of-mass frame.

The generalisation of the $R$-matrix to the case of proton collisions is straightforward. The full matrix is given by the weighted sum
of the various partonic channels, weighted by the luminosity functions. In the case of $\bar{t}t$ at Leading Order (LO) in QCD, in a proton collider, we have
\begin{align}
R(\hat{s}, \bm{k}) = \sum_I L^I(\hat{s}) R^I(\hat{s},\bm{k}) \,,
\label{eq:RmatrixWeightedLuminosity}
\end{align}
where $I=gg, \bar{q}q$, the two possible partonic channels. The spin density matrix is then defined simply as the normalised $R$-matrix, i.e.
\begin{equation}
\label{eq:rho}
\rho = \frac{R}{tr(R)}=\frac{\mathds{1}_2 \!\otimes\! \mathds{1}_2
+ B_i^+\sigma^i \!\otimes\! \mathds{1}_2 
+ B_i^- \mathds{1}_2 \!\otimes\! \sigma^i
+ C_{ij}\,\sigma^i \!\otimes\! \sigma^j}{4} .
\end{equation}
Quantum tomography has the objective of determining the various $B$ and $C$ coefficients in order to fully characterise the 
spin state of the system.
In order to obtain explicit values for the entanglement, we calculate the coefficients in the so-called helicity basis, which consists of an orthonormal basis in the center-of-mass frame
\begin{align}
\label{eq:helicity_basis}
\{\bm{k},\bm{n},\bm{r}\}: 
\,\,
\bm{r} = \frac{(\bm{p} - z\bm{k})}{\sqrt{1-z^2}},
\quad
\bm{n} = \bm{k}\times \bm{r},
\end{align}
where $\bm{p}$ and $\bm{k}$ are the unit vectors along the beam axis and top quark direction, and we define $z \equiv \bm{k}\cdot \bm{p} =\cos\theta$.

\section{Entanglement}

Whenever the quantum state of the system cannot be written as a convex combination of product states, i.e.
\begin{align}
\rho_{\rm ab}  = \sum_{k} p_k\, \rho^k_{\rm a}\otimes \rho^k_{\rm b} \, ,
\end{align}
the system is said to be entangled. This formal definition is more transparent when a quantitative measure of entanglement is given.
In the following, we quantify the degree of entanglement by defining a physical quantity called concurrence~\cite{Wootters:1997id}
\begin{align}
\label{eq:DeltaPeresHorodecki}
\Delta &\equiv  - C_{nn} + |C_{kk}  + C_{rr}| -1 > 0 \, ,\\
C[\rho]&=max(0, \frac{\Delta}{2}) \, .
\end{align}
When $C[\rho] > 0$, the system is said to be entangled and the case of $C[\rho]=1$ corresponds to quantum configurations of maximal entanglement.

\begin{figure}[t!]
\centering
  \includegraphics[scale=0.5]{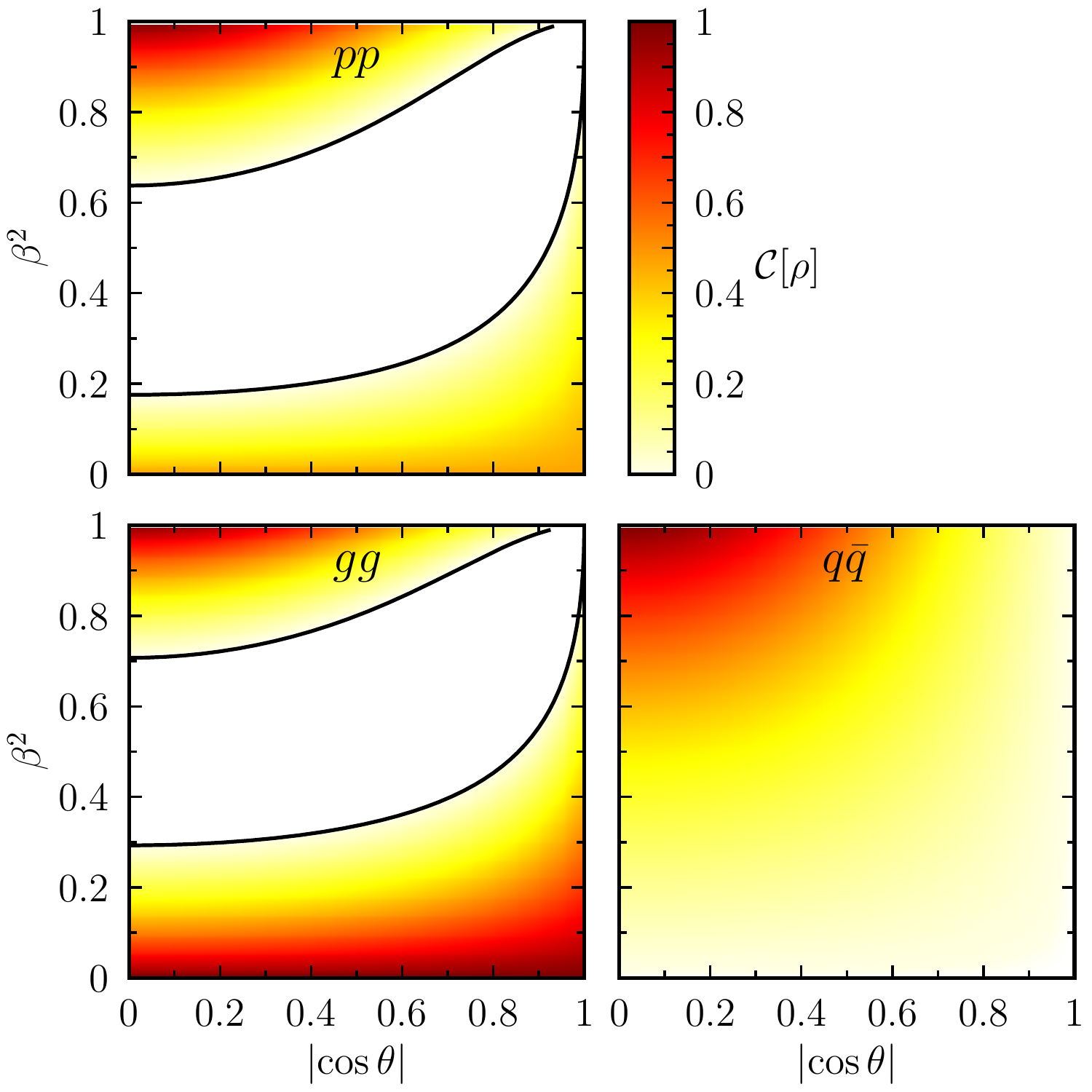}\,
  \caption{Standard Model contribution for the concurrence in the gluon-~(bottom left) and quark-initiated~(bottom right) channels, as well as in the full $pp$ collision~(top). The black lines indicate the boundaries of the entangled regions based on Eq.~\eqref{eq:DeltaPeresHorodecki}.
  }
  \label{fig:Concurrence_SM}
\end{figure}

In Fig.~\ref{fig:Concurrence_SM} we show the concurrence for top quark pair production in the SM as a function of $\beta^2$ and $z=|\cos{\theta}|$ and we identify the maximal entanglement regions.
In particular, for the $gg$ channel we have that at threshold and at high energy, the quantum state is given by
\begin{align}
\label{eq:QuantumStateSM}
    \rho_{gg}^{\text{SM}}(0,z) = |\Psi^-\rangle_{\bm n} \langle \Psi^-|_{\bm n},\quad
    \rho_{gg}^{\text{SM}}(1,0) = |\Psi^+\rangle_{\bm n} \langle \Psi^+|_{\bm n},
\end{align}
where
\begin{align}
    |\Phi^\pm \rangle_{\bm n} = \frac{|\!\uparrow \uparrow\rangle_{\bm{n}} \!\pm
    |\!\downarrow\downarrow\rangle_{\bm{n}} }{\sqrt{2}},
    \quad
    |\Psi^\pm \rangle_{\bm n} = \frac{|\!\uparrow \downarrow\rangle_{\bm{n}}  \!\pm|\!\downarrow \uparrow \rangle_{\bm{n}}}{\sqrt{2}} \, .
\end{align}
We therefore see that at threshold the top quark pair is in a spin-0 singlet state, while at high energy in a triplet state, both maximally entangled.
In the case of $\bar{q}q$ channel, we instead see that at threshold no entanglement is present, while at high energy the same maximally entangled triplet state is found.

\section{SMEFT effects at threshold}

To study the impact of higher-dimensional operators on the entanglement in top quark pair production within the SMEFT, we use a slightly modified version of the Warsaw basis~\cite{Grzadkowski:2010es},
\begin{align}
\LL_{\rm SMEFT} = \LL_{\rm SM} +  \frac{1}{\Lambda^2}\sum_i c_i \OO_i \,,
\end{align}
where we restrict ourselves to $CP$-even operators at dimension-six.

Here we show effects of the higher dimensional operators to the quantum state at threshold, see Ref.~\cite{Aoude:2022imd} for a more in depth analysis.
Regarding the $gg$ partonic channel, we find that NP effects induce a triplet state of spin 1, i.e.
\begin{align}
\rho_{gg}^{\rm EFT}(0,z) = p_{gg} |\Psi^+\rangle_{\bm p}\langle \Psi^+|_{\bm p}
+
(1-p_{gg})|\Psi^-\rangle_{\bm p}\langle \Psi^-|_{\bm p} \,.
\end{align} 
Note that here the spins are defined with respect to the beam direction $\bm{p}$. 
The probability of being in a triplet state is given by
$p_{gg} = 72 m_t^2 (3\sqrt{2} m_t \, c_G + v \, c_{tG})^2/7\Lambda^4 \, ,$
which shows that no linear effects are present and only the squares contribute. As a consequence, the system is not anymore in a 
pure maximally entangled state, and the entanglement is therefore reduced.
In the case of $\bar{q}q$ we instead find that the quantum state is given by
\begin{equation}
    \rho_{q\bar{q}}^{\rm EFT}(0,z) = p_{q \bar q} \ket{\uparrow \uparrow}_{\bm{p}}\bra{\uparrow \uparrow}_{\bm{p}} + (1-p_{q \bar q})\ket{\downarrow \downarrow}_{\bm{p}}\bra{\downarrow \downarrow}_{\bm{p}} \, ,
\end{equation}
where $p_{q\bar q} = \frac{1}{2}-4\frac{c_{VA}^{(8),u}}{\Lambda^2}$, with $c_{VA}^{(8),u} = (-c_{Qq}^{(8,1)} -c_{Qq}^{(8,3)} + c_{tu}^{(8)} - c_{tq}^{(8)} + c_{Qu}^{(8)})/4$.
The spoiling of the symmetry is due to P-violating interactions induced by dimension-six operators but is also present if electroweak corrections are taken into account. However, this state is still a separable one and therefore, as in the case of the SM, no entanglement is present.

\begin{figure}
  \centering
  \includegraphics[width=.5\linewidth]{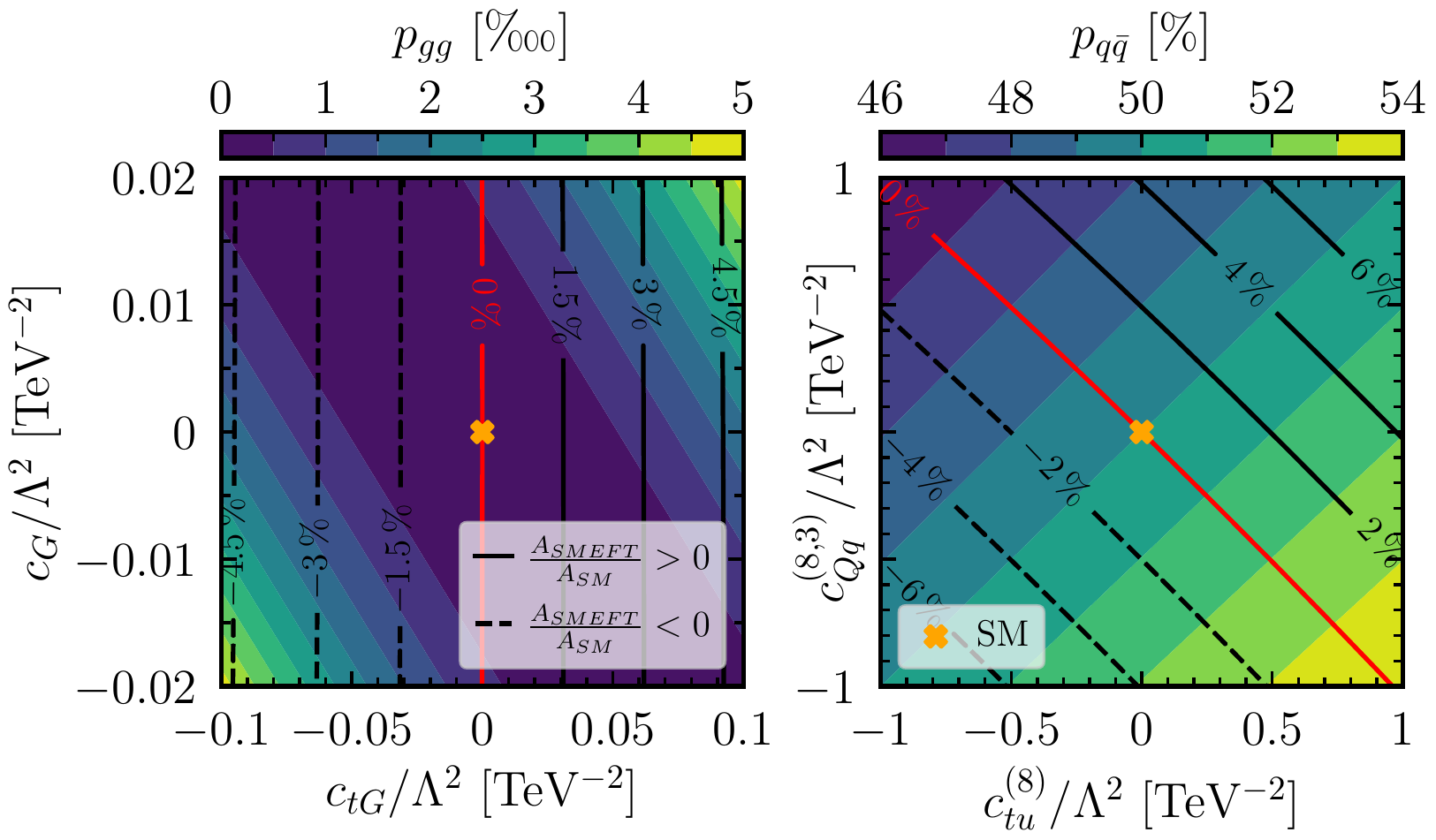}
  \caption{Probability to produce a triplet~(left) or both-spin-up~(right) state at threshold ($\beta=0$) in the $gg$ or $q\bar{q}$ channel, respectively. The contour lines indicate the relative corrections of the EFT to the scattering amplitude.}
    \label{fig:prob_threshold}
 \end{figure}

In Fig.~\ref{fig:prob_threshold} we show contour plots of the probabilities $p_{gg}$ and $p_{q\bar q}$. In the case of the quark initiated channel, we choose $\OO_{tu}^{(8)}$ and $\OO_{Qq}^{(8,3)}$ as a pair of representative four-fermion operators. In addition to the probabilities, we also plot contours of the relative EFT effects on the scattering amplitude, in order to highlight the complementarity of the two observables, which are clearly probing different directions in the parameter space.

\section{Conclusions}

In this work we have studied the effects of heavy NP to the entanglement patterns in top quark pair production at the LHC. In particular, we
focused on the maximally entangled regions according to the SM, i.e. the production at threshold and at high energy.
We found that, the maximally entangled regions are indeed affected and in the presence of
SMEFT operators the degree of entanglement is in general reduced.
With entanglement being at the core of quantum mechanics, one might hope that it will provide fundamental information on the structure of the effective field theory as much as unitarity, analyticity and positivity do on general properties of the scattering amplitudes. On a more practical level, 
a natural question worth posing is how much these new observables will help in better constraining top-quark SMEFT operators in global fits, also in comparison with usual spin correlation measurements.



\bibliographystyle{utphys}
\bibliography{refs}

\end{document}